\documentclass[preprint,journal]{vgtc}            

\onlineid{1329}

\vgtccategory{Research}

\title{FSLens: A Visual Analytics Approach to Evaluating and Optimizing the Spatial Layout of Fire Stations}

\author{%
\authororcid{Longfei Chen}{0009-0002-4596-8093}, \authororcid{He Wang}{0009-0003-2550-6139}, \authororcid{Yang Ouyang}{0009-0000-5841-7659}, Yang Zhou, Naiyu Wang, and \authororcid{Quan Li}{0000-0003-2249-0728}
}

\authorfooter{
  \item Longfei Chen, Yang Ouyang, He Wang, and Quan Li are with the School of Information Science and Technology, ShanghaiTech University, and Shanghai Engineering Research Center of Intelligent Vision and Imaging, China. Quan Li is the corresponding author. E-mail: \{chenlf,ouyy,wanghe1,liquan\}@shanghaitech.edu.cn.
  \item Yang Zhou and Naiyu Wang are with the College of Civil Engineering and Architecture, Zhejiang University. E-mail: \{yang-zhou,naiyuwang\}@zju.edu.cn.
}

\abstract{
  The provision of fire services plays a vital role in ensuring the safety of residents' lives and property. The spatial layout of fire stations is closely linked to the efficiency of fire rescue operations. Traditional approaches have primarily relied on mathematical planning models to generate appropriate layouts by summarizing relevant evaluation criteria. However, this optimization process presents significant challenges due to the extensive decision space, inherent conflicts among criteria, and decision-makers' preferences. To address these challenges, we propose \textit{FSLens}, an interactive visual analytics system that enables in-depth evaluation and rational optimization of fire station layout. Our approach integrates fire records and correlation features to reveal fire occurrence patterns and influencing factors using spatiotemporal sequence forecasting. We design an interactive visualization method to explore areas within the city that are potentially under-resourced for fire service based on the fire distribution and existing fire station layout. Moreover, we develop a collaborative human-computer multi-criteria decision model that generates multiple candidate solutions for optimizing firefighting resources within these areas. We simulate and compare the impact of different solutions on the original layout through well-designed visualizations, providing decision-makers with the most satisfactory solution. We demonstrate the effectiveness of our approach through one case study with real-world datasets. The feedback from domain experts indicates that our system helps them to better identify and improve potential gaps in the current fire station layout.
  
}


\teaser{
  \centering
  \includegraphics[width=\linewidth]{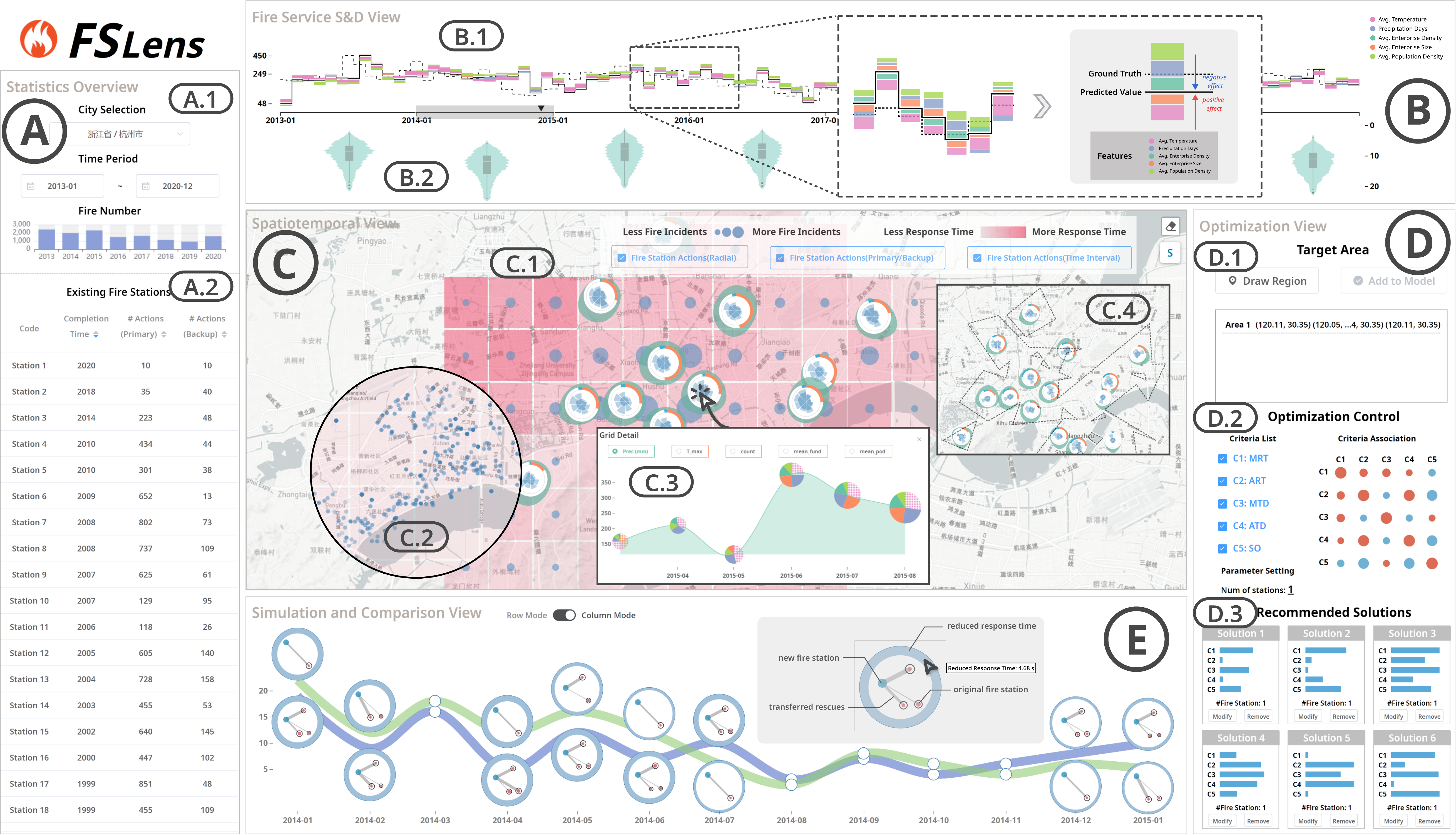}
  \vspace{-6mm}
  \caption{System Overview of \textit{FSLens}. (A) The Statistics Overview displays the statistical information of historical fires and fire stations. (B) The Fire Service S\&D View serves as a tool for experts to comprehend the fluctuations in the supply and demand of firefighting resources over time. (C) The Spatiotemporal View employs a map-based exploration method to exhibit the spatial distribution of fire incidents and the spatial layout of fire stations. (D) The Optimization View offers a set of interactions to support the user in generating multiple optimization scenarios for consideration. (E) The Simulation and Comparison View aids in the assessment of the effects of the optimization solutions on the original layout and offers a comparative evaluation of the efficacy among solutions.}
  \label{fig:teaser}
}

\graphicspath{{figs/}{figures/}{pictures/}{images/}{./}} 

\usepackage{tabu}                      
\usepackage{booktabs}                  
\usepackage{lipsum}                    
\usepackage{mwe}                       
\usepackage{balance}
\usepackage{mathptmx}                  

\usepackage{paralist}

\usepackage{amsmath}
\usepackage{amssymb}
\usepackage{mathrsfs}
\DeclareMathAlphabet{\mathcal}{OMS}{cmsy}{m}{n}

\usepackage{algorithm}
\usepackage{algorithmic}

\begin{document}


\firstsection{Introduction}

\maketitle
\par Fire stations play a crucial role in mitigating fires, which pose a threat to human life, property, and the environment~\cite{pellegrini2018fire,luo2019effect,zhang2019new}. The World Fire Statistics reported by CTIF\footnote{CTIF was founded in $1900$ in Paris for encouraging and promoting cooperation among firefighters and experts in Fire \& Rescue throughout the world.} indicate that approximately four million fires (affecting roughly $3.3$ billion individuals) occurred in $48$ countries worldwide in $2020$, resulting in around $20,700$ fatalities. Uncontrolled fires can spread quickly and pose an even greater hazard, highlighting the importance of efficiency in fire rescue operations.

\par The spatial layout of fire stations is closely associated with rescue efficiency~\cite{kc2018spatial,liu2020applying}, with an optimal layout significantly reducing response time\footnote{The response time is the duration elapsed between the instant the fire alarm is received and the time when the fire department arrives at the incident location.} and minimizing the harm resulting from a fire. There are many traditional studies focused on evaluating and optimizing the spatial layout of fire stations. The most common approaches~\cite{toregas1971location,church1974maximal,daskin1983maximum,batta1989maximal,weaver1985median} are usually based on the Location Set Covering Problem (LSCP) and the Maximum Covering Location Problem (MCLP). These methods determine the optimal fire station layout by minimizing the average travel distance or travel time to deliver fire services. Due to the complexity of real-life factors affecting fire station layout, more logical methods that consider the influence of multiple criteria~\cite{badri1998multi,schreuder1981application,monarchi1977simulation,schilling1980some,hogg1968siting} are widely used. The criteria considered in previous studies were generally related to rescue capability, construction costs, technology, and politics. Over the past few years, there has been an increasing trend of combining planning models with Geographic Information System (GIS) technology~\cite{chaudhary2016application,dong2018study,csen2011gis,uddin2020decision,erden2010multi} to analyze fire station layouts and achieve more precise evaluation and optimization.

\par Despite the initial effectiveness demonstrated by the aforementioned approaches in certain scenarios, they still encounter the following three challenges. \textbf{(1) Interpretability is needed to support decision-making.} In traditional approaches, the spatial layout of fire stations is influenced by multiple criteria, and the solution to such a multi-criteria optimization problem often lacks interpretability. For example, it is not easy to know which criterion plays a more important role in layout generation. Given that fire department planners typically lack a background in computer science, the limited interpretability of the solution is likely to impede their ability to comprehend and place trust in the recommendations provided by the solution. \textbf{(2) Inherent conflicts in multi-criteria planning.} In reality, fire station siting decisions often depend on various criteria, such as traffic conditions, construction costs, political indicators, and water supply. According to the literature~\cite{badri1998multi}, there may be potential conflicts among these criteria, which can lead to an optimization problem without a unique optimal solution. In the case of conflicting criteria, one criterion is optimized at the cost of some other criteria. Therefore, how to balance these criteria is a critical and challenging task that usually requires human analysis on a case-by-case basis. \textbf{(3) ``Tug-of-war'' between the whole and parts.} The preceding method entails the uniform placement of multiple fire stations across a broad geographical area, typically with urban centers. However, conventional algorithms may encounter limitations with respect to solution satisfaction when dealing with large-scale spatial data and complex site selection objectives~\cite{church1999location}. According to our collaborative experts, the motivation to optimize the spatial layout of fire stations often arises from identifying an ``imbalance between supply and demand'' of firefighting resources in certain locations, where specific areas may be situated too far from fire stations to receive timely assistance. While addressing imbalances at a local level is more practical and efficient, pinpointing areas of local imbalance and conducting further analysis remain a challenge. Although certain methods~\cite{dong2018study} can identify potential regions automatically, conducting a fine-grained analysis still demands manual effort, which can be labor-intensive and inflexible.

\par The assessment and optimization of the spatial layout of fire stations belong to the category of service facility planning. In this regard, visual analytics (VA) approaches have emerged as a critical tool in the investigation of such challenges. Prior research has predominantly focused on facility location selection problems, such as those pertaining to warehouses~\cite{li2020warehouse}, rental housing~\cite{weng2018homefinder}, billboards~\cite{liu2016smartadp}, and retail stores~\cite{karamshuk2013geo}. However, the potential of VA to optimize extant fire station layouts to accommodate evolving firefighting needs remains an unexplored domain. In contrast to previous studies, the selection of a new fire station site necessitates \textbf{\textit{a comprehensive evaluation of the existing layout}}, as well as \textbf{\textit{an appraisal of the impact of the new station on the existing infrastructure}}. This study aims to address the aforementioned challenges by closely integrating theoretical research and practical domain expertise in three key areas. \textbf{First}, we employ an interpretable spatiotemporal forecasting model that utilizes fire records and related features to uncover patterns and influencing factors in fire occurrences. \textbf{Second}, we develop an interactive visualization-driven pipeline that leverages fire distribution and existing fire station layouts to identify areas within cities with inadequate firefighting resources. \textbf{Third}, we propose a collaborative human-computer multi-criteria decision model that generates solutions through an algorithm while demonstrating the correlation between criteria. Furthermore, we employ elaborate visualizations to simulate and compare the impact of different solutions on the original fire station layout, providing decision-makers with optimal solutions. The primary contributions of this study are summarized as follows: 1) Systematic characterization of domain experts' requirements in evaluating and optimizing fire station layout, 2) Development of interactive visualizations with novel features to support the investigation of potential issues and optimization options with the existing fire station layout, and 3) Demonstration of the effectiveness of the proposed approach through a case study and expert interviews.

\section{Related Work}
\subsection{Fire Station Layout and Urban Planning}
\par The provision of fire station services has traditionally been a crucial factor in safeguarding human life and property~\cite{murray2013optimising}. To evaluate the effectiveness of such services, established response time benchmarks are typically utilized, with the primary objective of mitigating the risk of loss of life and property. For instance, the National Fire Protection Association (NFPA)~\cite{nfpa20101710} has formulated guidelines recommending a maximum response time of $9$ minutes for $90\%$ of calls in urban areas. Comparable standards have also been proposed in other nations to guarantee a swift and efficient response to service requests~\cite{zhongming2009ministry}.

\par The assessment and improvement of fire station layout, primarily in relation to response time, have been extensively studied. Plane et al.~\cite{plane1977mathematical} focused on optimizing fire station placement in Denver, Colorado, based on response time as the coverage standard. Reilly et al.~\cite{reilly1985development} emphasized meeting potential demand and maximum response criteria. Various optimization criteria beyond response time have been explored. For example, Doeksen et al.~\cite{doeksen1976optimum} determined the optimal location of rural fire systems by considering minimum firefighting mileage and maximum protection of assets. Schreuder~\cite{schreuder1981application} employed a mathematical model to determine the minimum number of fire stations, their locations, and the required fire tenders to achieve town-wide response time targets. Badri et al.~\cite{badri1998multi} proposed a multi-objective modeling approach to address conflicting objectives in fire station placement, including response time, service overlap, and water availability. Additionally, modern techniques have integrated real-time traffic~\cite{liu2020dynamic,xu2021evaluating,yu2012optimization,liu2020applying}, GIS~\cite{murray2013optimising,chaudhary2016application,echeverria2018analysis,dong2018study,csen2011gis,erden2010multi}, and genetic algorithms~\cite{yang2007fuzzy} to enhance optimization accuracy and effectiveness. However, algorithmic approaches may not always yield satisfactory solutions, as human preferences and the comprehension of decision-makers without a computer science background play crucial roles. Therefore, we propose the development of an interactive visualization-oriented framework to assist decision-makers in effectively evaluating existing fire station layouts and creating satisfactory optimization solutions.

\subsection{Mathematical Programming Approach}

\par Past studies investigating fire station layouts have used mathematical programming models to achieve success. Earlier techniques, such as the maximal covering model~\cite{church1974maximal}, maximum expected covering model~\cite{daskin1983maximum}, extended version of the maximal expected covering model~\cite{batta1989maximal}, and vector median model~\cite{weaver1985median}, focused on minimizing the average travel distance or travel time to optimize the layout. However, these approaches only considered a single decision criterion, whereas real-world situations are often more complex~\cite{badri1998multi}. Other studies~\cite{schreuder1981application,monarchi1977simulation,schilling1980some,hogg1968siting} incorporated the rescue costs of the fire brigade and the economic damage caused by the fire as optimization objectives to model more complex decision spaces. Our study follows this trend and treats fire station layout optimization as a Multi-Criteria Decision Making (MCDM) process with criteria including response time, service overlap, etc.

\par Classic MCDM methods rank alternate solutions based on their similarity to the ideal solution~\cite{tzeng2011multiple}, but the ideal solution often depends on the task or may not exist due to conflicting criteria~\cite{aruldoss2013survey}. An improved approach is to involve human knowledge to make informed decisions~\cite{bergner2013paraglide,booshehrian2012vismon,carenini2004valuecharts}. Exploratory visualization is a common technique for user-centered decision-making. For example, \textit{LineUP}~\cite{gratzl2013lineup} enables interactive criteria combination and weight adjustment to explore the impact of different criteria combinations on ranking. \textit{WeightLifter}~\cite{pajer2016weightlifter} expands upon this by exploring the weight space in the MCDM process, enabling users to better understand how decisions are sensitive to weight changes. Inspired by previous research, we offer users several commonly used criteria and allow them to manipulate them in various ways. We also introduce visualization techniques to help users understand and compare the solutions provided by the model.

\subsection{Location-based Visualization}

\par Numerous location-based visualization studies have extensively explored the selection of facility locations in various domains such as \textit{warehouses}~\cite{li2020warehouse}, \textit{rental housing}~\cite{weng2018homefinder}, \textit{billboards}~\cite{liu2016smartadp}, \textit{retail stores}~\cite{karamshuk2013geo}, and \textit{city utility services}~\cite{zhang2014visual}. These studies typically rely on comparing a set of potential locations to identify the most suitable one, while neglecting the interplay between facilities when selecting a location. In our study, the impact of each additional fire station on the original layout needs to be considered~\cite{murray2013optimising}. Additionally, the influence of the new fire station on future fire service delivery is a crucial issue to be addressed. Thus, the conventional VA techniques used in prior studies are not well suited for our scenario.

\par Location-based visualization usually involves spatiotemporal data with multiple attributes. For instance, \textit{WarehouseVis}~\cite{li2020warehouse} employed GPS trajectory data and warehouse data spanning a month, while \textit{SmartAdP}~\cite{liu2016smartadp} utilized taxi trajectory data and regional POI (i.e., point of interest) data within a specific time frame. Effective analysis of spatiotemporal data often necessitates well-designed visualization methods that can integrate the temporal and spatial features of the data~\cite{andrienko2003exploratory}. Generally, heat maps~\cite{hilton2011saferoadmaps,maciejewski2009visual,mehler2006spatial}, choropleth maps~\cite{stein2016game} and glyph-based visualizations~\cite{10.2312:conf:EG2013:stars:039-063} are used to depict the spatial distribution of location data. These techniques are intended to assist users in perceiving and comprehending the spatial characteristics of the data. Additionally, various timeline-based methods are equally powerful in presenting temporal features~\cite{aigner2011visualization}. In this work, we present enhanced timeline and map visualization techniques that merge the two to represent the spatiotemporal properties associated with fire stations.

\section{Observational Study}

\subsection{Experts' Conventional Practice and Bottlenecks}
\par The assessment and optimization of fire station layout is a multidisciplinary research issue that necessitates input from numerous domains, such as fire safety, mathematical programming, and urban computing. To gain insight into the practical approach taken by domain experts, we collaborated with a team of experts from a fire department and a partner university. This team included a fire service manager (E1, Male, Age: $31$) as well as two researchers (E2, Female, Age: $37$; E3, Male, Age: $26$) who specialize in the interdisciplinary realm of public service and urban computing technology.

\begin{figure*}[h]
    \centering
        \vspace{-3mm}
    \includegraphics[width=\textwidth]{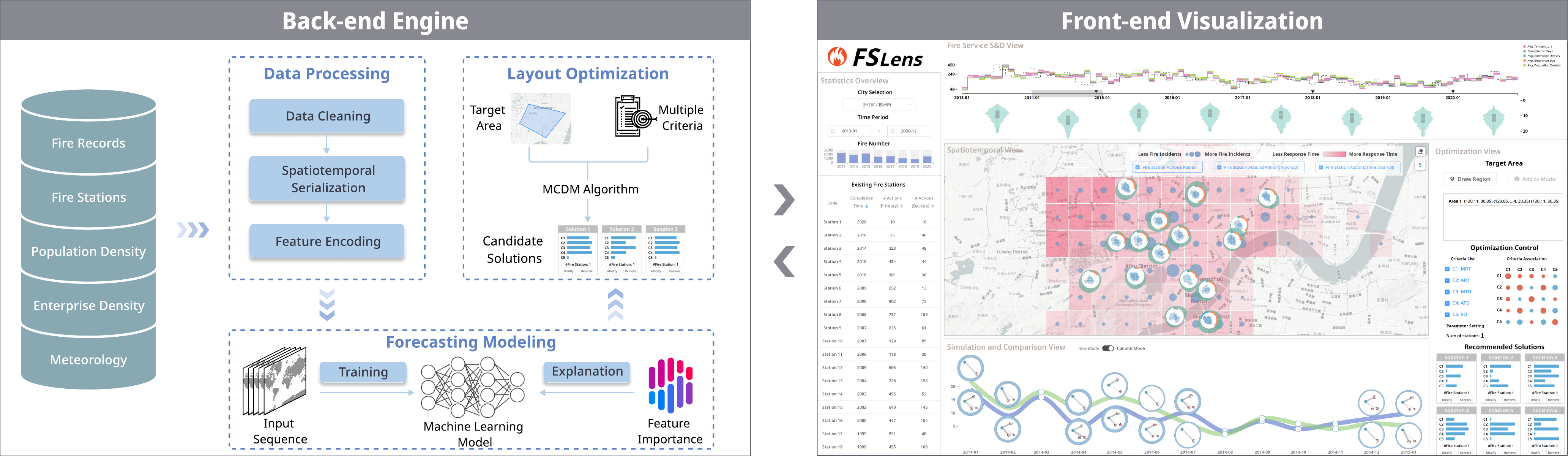}
    \vspace{-6mm}
    \caption{The system architecture consists of two parts: a back-end engine and a front-end visualization. The back-end engine comprises three modules: data processing for converting raw data into spatiotemporal sequences, forecasting for training a machine learning model, and layout optimization for suggesting fire station placements. The front-end visualization offers five views with advanced designs to facilitate effective analysis.} 
    \label{fig:pipeline}
    \vspace{-6mm}
\end{figure*}

\par The team of experts shared with us the practical approach utilized by the fire department for analyzing fire station layout. E1 noted that ``\textit{the traditional and most commonly used approach is a human-centric process, with decisions based on statistical information about fires and field research}''. E2 added that this approach is both costly and labor-intensive. To alleviate the workload of decision-makers, the fire department and the agency where E2 works developed an internal information system that integrates road network data, historical fire data, and existing fire station data with a map. Using this system, experts can gain a preliminary understanding of the distribution of fires and fire stations and identify areas with insufficient fire services resources by examining the coverage of existing fire stations and the frequency of fires. Once areas for optimization have been identified, offline research is conducted to determine the optimal strategy. E1 noted that ``\textit{the relocation of existing fire stations is challenging, and therefore, the typical approach is to select additional fire stations}''. Despite these efforts, the existing approach has limitations in mining data and informing decision-making, and it involves significant manual efforts, making it relatively inefficient overall. Additionally, the experts expressed concerns about integrating urban computing techniques and fire station layout analysis. E2 and E3 attempted to use mathematical planning models to aid in decision-making, but decision-makers found them challenging to comprehend and trust without relevant expertise. E3 further stated that ``\textit{it is crucial for decision-makers to understand the relationship between multiple criteria when analyzing the spatial layout of fire stations}''. 

\subsection{Experts' Needs and Expectations}

\par To identify primary concerns related to the assessment and optimization of fire station layout, as well as the possible impediments to making informed decisions, we interviewed E1--E3. The obtained interview feedback enabled us to distill the ensuing requirements.

\par \textbf{R.1 Reveal patterns and underlying factors contributing to fire incidents.} The fire station layout is heavily impacted by the incidence of fires. Analysis of fire patterns enables the identification of regions at high risk of fires, facilitating the allocation of fire stations in a manner that maximizes the utilization of fire service resources, thereby reducing costs. In addition, E2 expressed an interest in the identification of factors that contribute to the occurrence of fires, which could be used to anticipate potential fires and plan for the effective deployment of fire services in a forward-looking manner.

\par \textbf{R.2 Assess the rescue capability of the current fire station layout.} E1 emphasized the significance of being able to promptly and effectively respond to fires in all parts of the city as a critical aspect of the proper layout of fire stations. The rescue capacity of fire stations plays a crucial role in saving lives and mitigating property damage during fire incidents. Hence, conducting a comprehensive multi-criteria assessment of fire station rescue capability is imperative and a prerequisite for any subsequent optimization efforts. 

\par \textbf{R.3 Identify and explore regions where fire service resources are scarce.} Consistent with their prior practices, the experts expressed an interest in identifying regions with limited access to fire service resources. Additionally, they expressed a desire to conduct a more in-depth analysis of these areas based on meteorological, demographic, and other fire-related data. The availability of supplementary information can facilitate a more comprehensive understanding of the reasons for the inadequacy of firefighting resources in these areas and serve as a valuable reference for the development of optimization strategies.

\par \textbf{R.4 Provide viable solutions for optimizing the fire station layout.} During the interview, E3 noted that layout optimization algorithms featuring multiple criteria, commonly employed in research, may prove challenging for decision-makers to comprehend in practice. Conversely, the manual generation of optimization strategies in the absence of computational support can be both time-consuming and yield suboptimal outcomes. Accordingly, there is a pressing need for an  approach that can facilitate user comprehension of the evaluation criteria and recommended solutions derived from the optimization model.

\par \textbf{R.5 Model the effect of additional fire stations on the overall layout.} The incorporation of (a) new fire station(s) into the existing layout is anticipated to exert a notable influence on the forthcoming firefighting operations across the region. This challenge has the potential to enhance the efficiency of fire rescue activities within the surrounding area while simultaneously alleviating the workload of other fire stations. E1 emphasized the importance of scrutinizing the impact of the new fire station on neighboring stations to ensure optimal resource utilization and effective coordination of the overall firefighting system. Particularly, the experts expressed a desire for our approach to facilitating ``what-if'' analyses and furnishing insights into future scenarios.

\section{System Overview}
\par We introduce \textit{FSLens}, an interactive VA system that aids in evaluating and optimizing the spatial arrangement of fire stations. The system architecture (\cref{fig:pipeline}) consists of two components: a back-end engine and a front-end visualization. Within the back-end engine, three modules are integrated: \textit{data processing}, \textit{forecasting modeling}, and \textit{layout optimization}. In the data processing module, the initial raw data is divided into temporal and spatial series to transform it into spatiotemporal sequences. The forecasting module employs machine learning techniques to train a model on the input data and predict future fire incidents while improving interpretability by identifying important input features. The layout optimization module employs multiple criteria to suggest potential solutions for placing new fire stations in a specified area. The front-end visualization comprises five views, each incorporating sophisticated designs to facilitate effective analysis.

\section{Back-end Engine}

\subsection{Data Description and Processing}
\par In this study, we collaborate with experts from the fire department who provided us with de-identified data related to fires and fire stations. The fire data consists of records documenting fires that took place in Hangzhou's primary urban areas from $2006$ to $2022$, including information such as accident addresses, alarm times, response times, and arrival times. The fire station data provides an overview of all the fire stations in Hangzhou, including details such as geographical locations, service durations, and staffing levels. Overall, these datasets encompass $29,242$ fire records and $42$ fire stations spread across the city. Additionally, based on recommendations from domain experts, we acquire publicly available supplementary datasets, including population density data\footnote{Population density measures the number of people per unit of land area. (data source: https://hub.worldpop.org)}, enterprise density data\footnote{Enterprise density refers to the average number of enterprises per unit of land area. (data source: http://www.gsxt.gov.cn)}, and meteorological data\footnote{Meteorological data comprises information about weather conditions. (data source: https://rp5.ru/Weather\_in\_the\_world)}. The experts believe that these additional datasets can enhance our understanding of the factors influencing fires.

\par To comprehend the patterns exhibited by fire incidents, we convert the pre-existing fire data into spatiotemporal sequence data. A multivariate spatiotemporal-series data forecasting model is then employed to capture the fundamental characteristics of the fire patterns. Notably, some pre-processing is necessary to generate spatiotemporal sequence data from the raw fire data. Specifically, we partition the urban region into several uniformly-sized grids based on geographical location and tail the occurrences of fires and other pertinent metrics (such as population density, business density, and temperature) within each grid at fixed time intervals. The grid size and time interval are determined through consultations with domain experts to ensure that these parameters are meaningful in the context of real-world studies.

\subsection{Spatiotemporal Sequence Forecasting Modeling}

\par In this study, we employ machine learning models in conjunction with interpretability techniques to investigate the patterns and underlying factors that contribute to fire incidents. Since both the temporal and spatial aspects of fires are of interest, predicting fire occurrences can be considered a multivariate spatiotemporal sequence forecasting problem. To accomplish this, we divide the entire spatial area into equally sized grids of $M\times N$ and assign $C$ time-varying features to each grid. As a result, the overall situation at any given timestamp can be represented as a tensor $\mathcal{X}\in \mathbb{R}^{C\times M\times N}$, where $\mathbb{R}$ signifies the feature range. We then calculate the overall situation at fixed time intervals using statistical methods to generate a sequence of tensors $\hat{\mathcal{X}}_1, \hat{\mathcal{X}}_2, ..., \hat{\mathcal{X}}_t$. Using the aforementioned notations, we can represent the spatiotemporal sequence forecasting problem as follows.
\begin{equation}
    \mathop{\arg\max}\limits_{\mathcal{X}_{t+1}, ..., \mathcal{X}_{t+K}} p(\mathcal{X}_{t+1}, ..., \mathcal{X}_{t+K} | \hat{\mathcal{X}}_{t-T+1}, \hat{\mathcal{X}}_{t-T+2}, ..., \hat{\mathcal{X}}_{t}),
\end{equation}
where the notion $T$ and $K$ refer to the time intervals utilized by the model to forecast future values. Specifically, $T$ denotes the historical data interval encompassing past observations, whereas $K$ denotes the prediction interval comprising future time points.

\begin{figure}[h]
    \centering
     \vspace{-3mm}
    \includegraphics[width=0.5\textwidth]{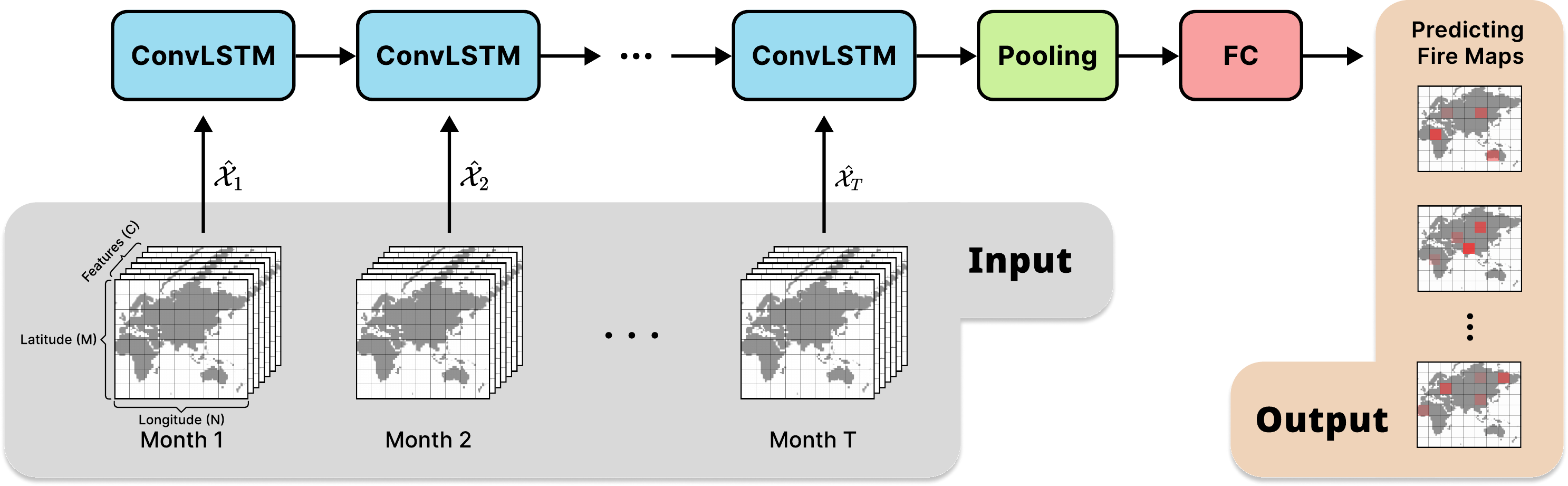}
    \vspace{-6mm}
    \caption{Structure of the \textit{ConvLSTM} network.}
    \label{fig:lstm}
    \vspace{-3mm}
\end{figure}

\par After discussion with domain experts, we select a grid size of $3\ km\times 3\ km$ resulting in a grid layout of $87\times 50$ cells and employ a temporal resolution of one month. The grid attributes are comprised of $5$ features, which include \textit{Avg. Temperature}, \textit{Precipitation Days}, \textit{Avg. Enterprise Density}, \textit{Avg. Enterprise Size}, and \textit{Avg. Population Density}. Drawing inspiration from~\cite{shi2015convolutional}, we adapt the \textit{ConvLSTM} (\cref{fig:lstm}), a widely used spatiotemporal sequence prediction model, to our fire prediction task. The effectiveness of \textit{ConvLSTM} model has been demonstrated in various domains, including but not limited to weather forecasting~\cite{shi2015convolutional,tekin2021spatio,tan2018forecast}, vegetation forecasting~\cite{robin2022learning}, video prediction~\cite{duzceker2021deepvideomvs} and traffic flow prediction~\cite{liu2020dynamic}. To enhance the precision of our predictions, we experiment with different network architectures by varying the number of layers and kernel sizes while utilizing the same training and test datasets. The experimental results, presented in \cref{table:nn_architectures}, are evaluated using the Root-Mean-Square Error (RMSE) metric. Based on these results, we determine that the most optimal performance is achieved using a three-layer network with all kernels sized $3\times3$. Each \textit{ConvLSTM} layer has hidden states of size $128$, $64$, and $64$, respectively. Subsequently, we conduct a comparative analysis between the \textit{ConvLSTM} model and two other spatiotemporal sequence prediction models: the Spatiotemporal Graph Convolutional Network (\textit{STGCN}) proposed by Yu et al.~\cite{yu2017spatio} and the XGBoost method~\cite{chen2016xgboost}. This analysis, detailed in~\cref{table:model_comparison}, employs three standard performance metrics: Mean Absolute Error (MAE), Mean Squared Error (MSE), and Root Mean Squared Error (RMSE). The results indicate that the ConvLSTM model outperforms the other models in terms of both MSE and RMSE metrics, showcasing its superior predictive accuracy. However, it is noteworthy that the \textit{ConvLSTM} model's performance in terms of MAE is similar to that of the \textit{STGCN model}, suggesting competitive predictive capabilities for the latter in this aspect.

\begin{table}[h]
\centering
\vspace{-3mm}
\caption{Comparative evaluation of different \textit{ConvLSTM} networks. ``(3x3)'' denotes the size of the input-to-state kernel, while ``[3x3, 256]'' designates a \textit{ConvLSTM} layer wherein the state-to-state kernel size is 3x3, and the number of hidden states is 256.}
\vspace{-3mm}
\label{table:nn_architectures}
\resizebox{\linewidth}{!}{
\begin{tabular}{lcc}
\toprule
\textbf{Network Architecture} & \textbf{Number of Parameters} & \textbf{RMSE} \\
\midrule
(3x3)--[3x3, 256]                       & $2,580,225$  & $1.025$ \\
(3x3)--[3x3, 128]--[3x3, 128]           & $1,880,449$  & $0.492$ \\
\textbf{(3x3)--[3x3, 128]--[3x3, 64]--[3x3, 64]} & $\textbf{1,705,633}$  & $\textbf{0.479}$ \\
(6x6)--[1x1, 128]--[1x1, 128]           & $2,200,449$  & $0.539$ \\
(6x6)--[1x1, 128]--[1x1, 64]--[1x1, 64] & $2,017,985$  & $0.536$ \\
\bottomrule
\end{tabular}}
\vspace{-3mm}
\end{table}

\par In order to investigate the factors that influence the performance of the \textit{ConvLSTM} model, we provide additional model explanations to facilitate comprehension of the model's performance and feature importance differences~\cite{xu2021mtseer}. Various methods have been employed to interpret neural network models, including activation visualization~\cite{olah2015understanding,karpathy2015visualizing}, attention mechanisms~\cite{bahdanau2014neural}, rule extraction~\cite{hailesilassie2016rule}, and SHapley Additive exPlanations (\textit{SHAP})~\cite{lundberg2017unified}. Compared to other interpretable techniques, \textit{SHAP} illuminates the correlation between input features and output values in a more intuitive manner that is closer to human intuition. Additionally, \textit{SHAP} values indicate the impact of features on each sample while displaying the positivity and negativity of their influence, aligning with our intention to design a subsequent visualization. Therefore, we select \textit{SHAP} as our technique to explain the fire prediction model.

\begin{table}[h]
\centering
\vspace{-3mm}
\caption{Comparison of different models in the fire prediction task.}
\vspace{-3mm}
\label{table:model_comparison}
\begin{tabular}{lccc}
\toprule
\textbf{Model} & \textbf{MAE} & \textbf{MSE} & \textbf{RMSE} \\
\midrule
\textit{ConvLSTM}       & $0.423$  & $\textbf{0.231}$ & $\textbf{0.479}$ \\
\textit{STGCN}          & $\textbf{0.375}$  & $0.454$ & $0.674$\\
\textit{XGBoost}        & $0.601$  & $0.675$ & $0.810$\\
\bottomrule
\end{tabular}
\vspace{-3mm}
\end{table}

\subsection{Optimizing the Fire Station Layout}
\label{sec:criteria}
\par Our system enables experts to interactively construct a model that supports the optimization of the current spatial layout of fire stations in areas identified as lacking firefighting resources. Typical optimization methods involve constructing new fire stations or relocating existing fire ones~\cite{murray2013optimising}. However, relocation approaches are often challenging to implement due to the associated costs. Experts concurred that building new fire stations is a more practical and efficient approach to optimization, as well as being an MCDM process. Accordingly, we adopt this viewpoint and propose a visualization-oriented optimization model that not only computes reasonable locations based on multiple criteria but also unveils the interrelationships between the criteria.

\par Drawing on literature research~\cite{badri1998multi} and expert interviews, we have formulated the following set of criteria: 1) \underline{\textit{Average Response Time (ART)}}: Minimizing the average time traveled from (a) station(s) to fire scenes. 2) \underline{\textit{Maximum Response Time (MRT)}}:  Minimizing the maximum time traveled from (a) station(s) to fire scenes. 3) \underline{\textit{Average Travel Distance (ATD)}}: Minimizing the average distance traveled from (a) station(s) to fire scenes. 4) \underline{\textit{Maximum Travel Distance (MTD)}}: Minimizing the maximum distance traveled from (a) station(s) to fire scenes. 5) \underline{\textit{Service Overlap (SO)}}: Minimizing service overlaps between (a) new fire station(s) and existing fire stations. We then represent the MCDM problem as $\mathop{\min}\ F(x) = \{f_1(x), ..., f_m(x)\}^T \quad \text{s.t.}\quad x\in\Omega,$ where the notation $x=(x_1,x_2,...,x_k)$ denotes a set of additional fire stations (i.e., decision variables), $\Omega$ denotes the target area (i.e., decision space), and $F:\Omega\rightarrow \mathbb{R}^m$ consists of $m$ objective functions, which is a subset of \{\textit{ART}, \textit{MRT}, \textit{ATD}, \textit{MTD}, \textit{SO}\}.

\par During implementation, an expert had the option to interactively specify the desired number of new fire stations, delineate a target area, and select objective functions via the front-end visualization interface (introduced later). The back-end algorithm subsequently calculates and optimizes the objective functions utilizing data on the fire conditions within the target area. Ultimately, the algorithm furnishes the corresponding solution(s) to the expert.

\par Determining the location of $k$ additional fire stations is a computationally challenging task owing to the interactions between objectives. Specifically, this problem is classified as NP-hard due to the significant complexity involved. As the value of $k$ increases, the computational requirements for identifying an optimal solution escalate exponentially, rendering it infeasible to obtain a polynomial time solution. Therefore, we propose the adoption of a classical heuristic genetic algorithm, \textit{NSGA-\uppercase\expandafter{\romannumeral2}}~\cite{deb2002fast}, to strike a balance between efficiency and effectiveness. The algorithm \textit{NSGA-\uppercase\expandafter{\romannumeral2}} for multi-criteria optimization can be succinctly described as follows. \textbf{(1) Initialization.} An initial population of solutions is randomly generated and their fitness is evaluated based on the optimization objectives. \textbf{(2) Non-dominated sorting.} The population is sorted into non-dominated fronts according to their dominance relationships. \textbf{(3) Crowding distance computation.} The crowding distance for each solution in each front is computed to indicate the density of solutions in the objective space. \textbf{(4) Selection.} Binary tournament selection~\cite{miller1995genetic} is used to create a mating pool for generating offspring, taking into account both front rank and crowding distance. \textbf{(5) Crossover and mutation.} Genetic operators are then applied to the mating pool, resulting in a new offspring population. \textbf{(6) Merge and create a new population.} The parent and offspring populations are merged, non-dominated sorting is performed, and the next generation's population is filled by selecting solutions based on front rank and crowding distance. \textbf{(7) Termination check.} A termination check is conducted to determine if a stopping condition is met. If not, the process is repeated from step 2 with the new population. If so, the non-dominated solutions in the final population are returned as the Pareto-optimal set. To explore the connections between objectives, we extract objective values for each solution in the Pareto-optimal set. These objectives are treated as discrete random variables, denoted as $Y={y_1, y_2, ..., y_n}$, where $y_i$ represents the objective value of the $i^{th}$ solution. To evaluate the interdependence among these objectives, we utilize the Pearson correlation coefficient~\cite{cohen2009pearson} to quantify the pairwise relationships between the variables.


    
    

\section{Front-end Visualization}
\par We extend familiar visual metaphors to meet the requirements and facilitate analysis for domain experts. Following the principle of ``Overview first, zoom and filter, then details-on-demand''~\cite{shneiderman2003eyes}, we develop five visualizations in our design, as depicted in~\cref{fig:teaser}. These visualizations enable easy inspection, evaluation, and optimization of the firefighting system in the region. The Statistics Overview displays statistical information on historical fires and fire stations. The Fire Service S\&D View tracks changes in the number and response time of fires over time, providing insights into factors influencing fire occurrence based on the forecasting model (\textbf{R.1}). The Spatiotemporal View projects fire incident locations and fire station locations onto map grids, enhancing understanding of fire distribution and station layout (\textbf{R.2 -- R.3}). The Optimization View offers location recommendations for new fire station construction in a specific area, considering optional criteria (\textbf{R.4}). The Simulation and Comparison View facilitates the assessment of optimization solutions' effects on the original layout and allows for comparative evaluation among solutions (\textbf{R.5}).

\subsection{Statistics Overview}
\par The Statistics Overview presented in~\cref{fig:teaser}(A) provides valuable information on fire incidents and fire stations. The interface offers a control panel at the top, allowing users to load datasets for different cities and time periods. A bar chart in the middle of the view displays the number of fire incidents by year. Additionally, a table is employed to present statistical data on the fire stations, including their identification codes, completion times, and the number of actions. Notably, the number of actions is a cumulative count, determined according to the selected time period in the control panel. Furthermore, as fire stations may be categorized as either ``primary'' or ``backup'' in terms of their respective roles in fire rescue actions, the table includes a count of the number of fire stations in each category.

\subsection{Fire Service Supply \& Demand View}
\par The Fire Service Supply \& Demand (S\&D) View depicted in~\cref{fig:teaser}(B) serves as a tool for experts to comprehend the fluctuations in the supply and demand of firefighting resources over time. To begin with, it is important to note that ``supply and demand of firefighting resources'' refers to the equilibrium between the rescue resources provided by fire services and the demand for these resources by society. In this study, the frequency of fire incidents is employed to represent ``demand'', while the average response time of fire stations attending to incidents is used to represent ``supply''. These variables  are based on the benchmark criteria established by fire department experts in practice. E1 noted that ``\textit{an upsurge in fire incidents corresponds to an increased demand for fire services, and longer response times indicate inadequate resource allocation, highlighting a deficiency in service provision.}'' Utilizing the aforementioned parameters, we formulated two subplots that are delimited by the horizontal axis. 

\par \textbf{Step line chart with stacked bars indicates the fire number and causative factors in the forecasting model.} The upper subplot, as depicted in~\cref{fig:teaser}(B.1), illustrates the temporal fluctuations of the number of fires and their causative factors in the spatiotemporal forecasting model on a monthly basis. The model's predicted values are represented by a solid black line, while the actual number of fires is depicted by a dashed black line. Hence, by observing the difference between the predicted and ground-truth values at different timestamps, the user can evaluate the model's accuracy. Moreover, the importance of the five input features is computed, normalized, and exhibited in different colors. At each timestamp, stacked bars, representing the feature importance, are positioned along the vertical axis, and each bar in the stack corresponds to a feature. The bars corresponding to features with negative \textit{SHAP} values are positioned above the predicted values, implying that they have a negative impact and push the predicted values downwards. In contrast, features with positive \textit{SHAP} values are stacked below the predicted values, indicating that they have a positive effect and push the predicted values upwards~\cite{xu2021mtseer, zhang2022promotionlens}. It is noteworthy that since the utilized forecasting model is based on spatiotemporal sequence data, the resultant prediction outcomes and feature importance are also spatiotemporally distributed, as shown in \cref{fig:lstm}. The ground truth, predicted, and \textit{SHAP} values presented in this subplot are acquired by accumulating all grids at each timestamp to represent the variability of fires and influencing factors in the entire region.

\par \textbf{Violin plot indicates the response times distribution.} The lower subplot (\cref{fig:teaser}(B.2)) demonstrates the distribution of response times taken by fire stations to reach fire scenes. The fire records for each year are compiled, and their response times are calculated and displayed in a violin plot. The outer curve represents the actual distribution of response times, while the inner box plot represents the maximum, minimum, median, and quartile points of that distribution.

\par In this view, the demand for firefighting resources is proportional to the distance of the upper black dashed line from the horizontal axis, whereas the scarcity of such resources is inversely proportional to the distance of the lower violin plot from the same axis. Furthermore, the timeline of fire station construction is denoted on the horizontal axis to provide insight into the gradual development of the fire station layout.

\subsection{Spatiotemporal View}
\par The Spatiotemporal View, as depicted in~\cref{fig:teaser}(C), employs a map-based exploration method to exhibit the spatial distribution of fire incidents and the spatial layout of fire stations. 

\par \textbf{Hierarchical grid layer depicts the spatial distribution of fire incidents.} To avoid potential visual overlap and enable users to examine the data at different levels, we adopt a hierarchical design to depict the fire incidents. A grid layer is superimposed on the map, which varies in size depending on the level of zoom. At lower zoom levels (\cref{fig:teaser}(C.1)), the grid layer comprises several square glyphs of equal size. The central circle's area within each grid glyph corresponds to the number of fires within the covered area, and the glyph's color denotes the average response time for the fires in the region. As the zoom level increases (\cref{fig:teaser}(C.2)), the inner circle of the grid glyph will vanish and be replaced by small colored dots. Each dot represents a fire incident, with its color indicating the response time and its position signifying the location where the event occurred. Moreover, to facilitate the observation of temporal variations in the original features of each grid, we superimpose sector charts onto the area chart, as illustrated in \cref{fig:teaser}(C.3). Since the units of the original features are not uniform, users can switch between the five features' original values by clicking on the legend at the top of the tooltip. These features align with the Fire Service S\&D View and use the same color scheme. The sector charts compare feature importance at each timestamp, with their size indicating the summation of the absolute SHAP values for all features ($\Sigma_{all features} |SHAP|$) at the corresponding timestamp. Each sector's percentage in the sector chart denotes the corresponding feature's $\frac{|SHAP|}{\Sigma_{all features} |SHAP|}$, with a shading mask reflecting negative values.

\par
\begin{figure}[h]
    \centering
       \vspace{-6mm}
    \includegraphics[width=0.5\textwidth]{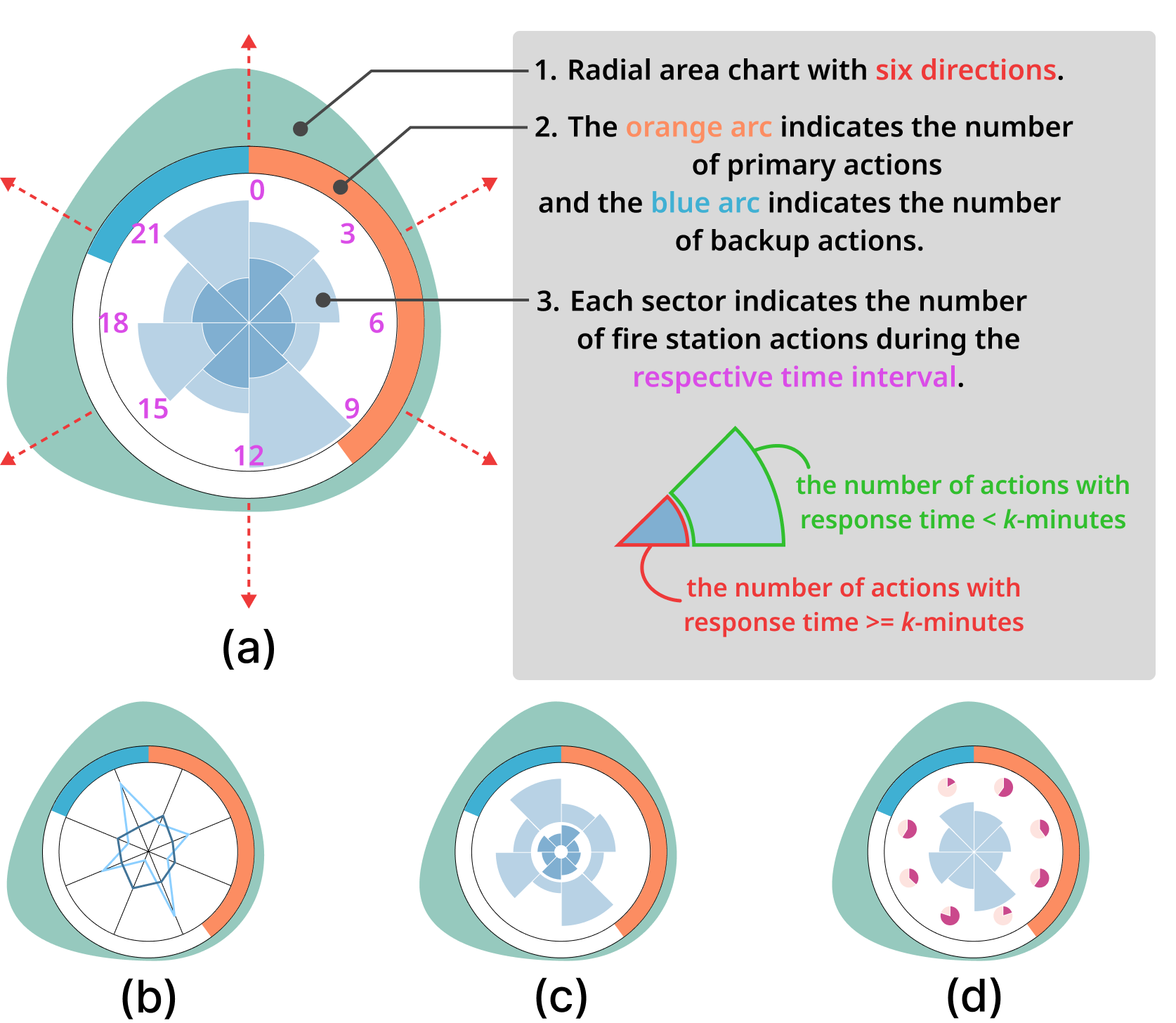}
    \vspace{-6mm}
    \caption{The stacked rose chart (a) was chosen for its ability to display and compare multiple data series while maintaining high information density and visualization space. The radar chart (b) was ineffective due to visual clutter and difficulties in comparing data series. The nested rose chart (c) allowed within-level comparisons but lacked total action information. Design (d) combined a base rose chart for total actions and an outer sector chart for response level percentages but hindered level comparisons and required significant space.}
    \label{fig:alter_glyph}
    \vspace{-6mm}
\end{figure}

 \textbf{Enhanced radial glyphs show information about the fire stations.} In order to enable users to obtain a concise overview of a fire station, we have developed a radial glyph. As shown in \cref{fig:alter_glyph}(a), this glyph comprises a radial area chart, which is situated around the glyph, and records the quantity of fire station actions in each of the six directions (\cref{fig:alter_glyph}(a.1)). The selection of the number of directions for our area map was made following consultations with domain experts. In practical applications, both the four cardinal directions (east, south, west, and north) and the eight ordinal directions (east, south, west, north, northeast, northwest, southeast, and southwest) are commonly employed. However, limiting the map to only four directions may result in an aesthetically less pleasing representation, while considering all eight directions could introduce a higher level of complexity to the map's shape. Consequently, after thorough deliberation with experts, we opted for a balanced compromise solution, employing six directions for our area map. The shape of the chart is determined by the relative location of the fire incidents and the station and reflects the fire station's service coverage. As shown in~\cref{fig:alter_glyph}(a.2), the length of the outer arc denotes the number of fire station actions, with the orange arc on the right indicating the ``primary'' role and the blue arc on the left indicating the ``backup'' role. Furthermore, as shown in~\cref{fig:alter_glyph}(a.3), the glyph incorporates a stacked rose chart that illustrates the frequency of fire station actions at different time intervals. The darker portion of each sector represents the number of actions that exceeded a response time threshold of no less than $k$-minutes, while the lighter portion corresponds to the number of actions that met the response time threshold of less than $k$-minutes. In accordance with NFPA standard~\cite{nfpa20101710}, the default value for $k$ is set to $9$ minutes and can be modified by users as per their actual requirements and preferences. To identify and investigate areas of potential scarcity in firefighting resources, we use black dashed boundaries (\cref{fig:teaser}(C.4)) to represent the reachable area of the current fire station layout based on the value of $k$. Users can adjust the value of $k$ by using the slider in the upper right corner of the map. Specifically, when closed boundaries are nested, access is restricted to the area enclosed between the innermost and outermost boundaries. To illustrate, within the largest closed boundary depicted in~\cref{fig:teaser}(C.4), multiple smaller polygons exist, and the regions encompassed by these smaller polygons are rendered inaccessible.

\par \textbf{Design Alternatives.} \cref{fig:alter_glyph} illustrates different designs examined and evaluated for conveying fire station profile information. Initially, a radar chart (\cref{fig:alter_glyph}(b)) was considered to display the number of actions at various response levels ($\geq k$-minutes and $<k$-minutes) during each time period. However, the radar chart's effectiveness was hindered by the need for polygon coverage analysis to compare data series, leading to visual clutter. In contrast, the nested rose chart (\cref{fig:alter_glyph}(c)) presented a more intuitive alternative. It featured two concentric circles representing different response levels, each with sectors for within-level comparisons. However, this design lacked a visual representation of total action numbers. \cref{fig:alter_glyph}(d) addressed this issue by utilizing a base rose chart to represent total actions, alongside an outer sector chart indicating the percentage of response levels in the corresponding time period. Nevertheless, this approach made it challenging to compare the same level across different time periods and required significant space, potentially obstructing other map layer information. Consequently, the stacked rose chart (\cref{fig:alter_glyph}(a)) was chosen as it enabled the simultaneous visual display and comparison of multiple data series, including total actions and response levels. This design offered higher information density within a limited space while preserving visualization space.

\subsection{Optimization View}
\par The Optimization View, illustrated in~\cref{fig:teaser}(D), offers a range of interactions to assist users in generating multiple optimization scenarios for consideration. The target area panel (\cref{fig:teaser}(D.1)) allows users to select specific areas on the map and add them to the optimization model, with each selected area being recorded. The optimization control panel (\cref{fig:teaser}(D.2)) presents a predefined list of optimization criteria (refer to \cref{sec:criteria} for definitions) and a parameter setting panel. By setting the criteria and parameters, the system activates the multi-criteria optimization model to generate multiple recommended solutions. The results provide correlations between different criteria, displayed in a matrix plot. The size of the points in the matrix plot represents the absolute value of the correlation coefficient, with red indicating positive values and blue indicating negative values. To represent the candidate solutions, we use boxes in the preview window (\cref{fig:teaser}(D.3)). Within each box, vertical bars indicate the optimization level of that solution for different criteria. Users can compare criteria vertically within the same solution and horizontally across different solutions. Moreover, clicking on a box visually represents the solution's location on the map as a pin. Users can refine the location by selecting the ``Modify'' button, allowing them to relocate the pin by dragging it. Similarly, the option to eliminate unsatisfactory solutions is provided through the ``Remove'' button, giving users the ability to discard undesirable selections.

\subsection{Simulation and Comparison View}
\par In the Simulation and Comparison View, as shown in~\cref{fig:teaser}(E), we address the experts' interest in understanding the impact of implementing the recommended solutions, particularly the effect of new fire station(s) on the existing layout. This view facilitates a ``what-if'' analysis of layout optimization solutions. We introduce an enhanced line chart where the horizontal axis represents time and the vertical axis represents the number of ``transferred rescues''. In our simulation, each fire is based on real data, representing an actual fire occurrence and rescue by a real fire station. A ``transferred rescue'' occurs when a fire previously handled by the original fire station is reassigned to a new simulated fire station. The reassignment is based on the simulation's travel time to the fire scene being shorter than the actual response time. Counting the number of ``transferred rescues'' helps measure the impact of the simulated new fire station on the original layout. Within the enhanced line chart, each line corresponds to a solution. The connection points on the line represent the impact of the new fire station on the original layout during the corresponding time period. Specifically, a blue node represents a new fire station, with its size indicating the number of rescues assigned to it by the simulation. A red node represents an existing fire station. The outer circle's size represents the number of real rescues before reassignment, while the inner circle's size represents the number of simulated rescues after reassignment. The edges between nodes represent the transfer relationship, with their thickness reflecting the number of rescues transferred from the original fire station to the new fire station. Additionally, the nodes' relative positions are determined by the actual geographical locations of the fire stations. When users add multiple solutions to this view, the system provides both row mode and column mode. In row mode, users can observe the impact of a solution on the original layout over time. In column mode, users can compare different solutions horizontally at the same timestamp.

\begin{figure*}[h]
\centering
\vspace{-3mm}
\includegraphics[width=\textwidth]{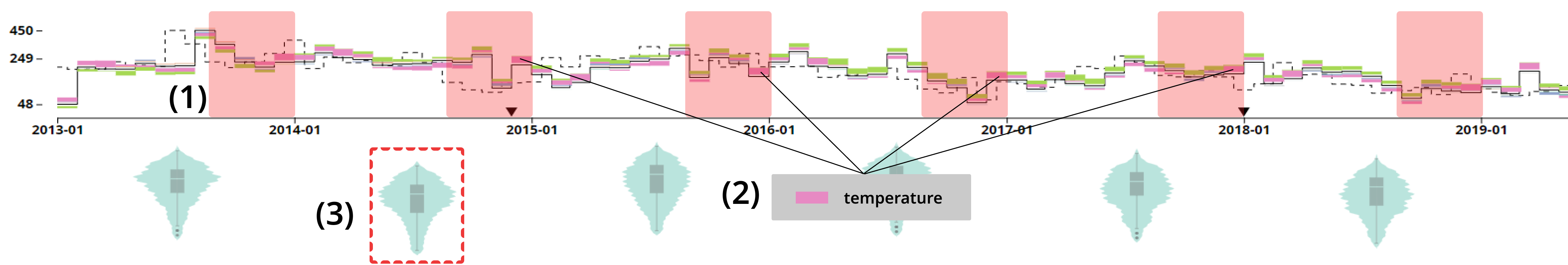}
\vspace{-6mm}
\caption{(1) The fires show a decreasing trend between Oct. and Jan. each year. (2) The temperature feature is more significant during this period and has a negative effect on the predicted results  (3) The response time for fires is significantly longer during the $2014$--$2015$ period than at other times.}
\label{fig:pattern}
\vspace{-6mm}
\end{figure*}

\section{Evaluation}

\subsection{Case Study}

\par The following are the activities performed by E1 -- E3 when analyzing the data related to fires and fire stations recorded in Hangzhou from $2013$ to $2020$ using the \textit{FSLens} system.

\par Upon loading the fire data recorded in Hangzhou from $2013$ to $2020$ into \textit{FSLens}, the experts proceeded to the Statistics Overview, which displays the fire and fire station information. The Fire Service S\&D View was then accessed by the experts, which presents the number of fires and response times in two separate subplots, top and bottom. As shown in \cref{fig:pattern}(1), the experts first observed a regular pattern in the occurrence of fires according to the seasons in the top subplot. E2 noted that ``\textit{the fires show a decreasing trend between October and January each year,}'' and attributed it to the temperature factor. Further analysis revealed that the temperature feature is more significant during this period and has a negative effect on the predicted results (\cref{fig:pattern}(2)). In the subplot below, the experts identified that the response time for fires was significantly longer during the $2014$--$2015$ period compared to other periods, which piqued their interest to explore the distribution of fires and fire stations during that time frame (\cref{fig:pattern}(3)). Following the brushing process, the Spatiotemporal View displayed relevant information. E3 observed that the fire station actions were predominantly recorded during the hours between $18:00$ and $24:00$. E3 posited that ``\textit{an increase in electricity usage at night could potentially lead to more fire instances.}''

\begin{figure}[h]
    \centering
    \includegraphics[width=0.5\textwidth]{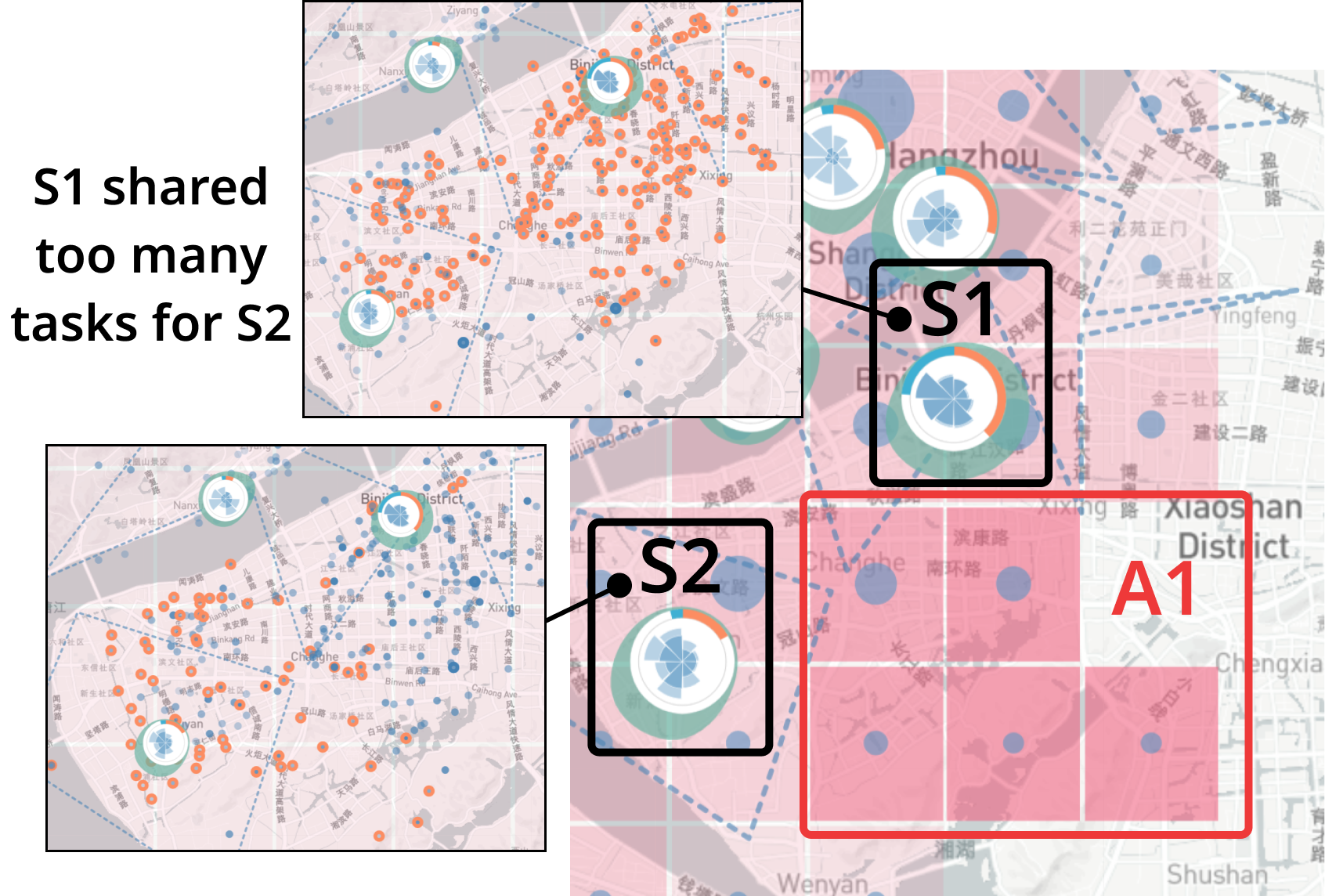}
    \vspace{-6mm}
    \caption{Experts found the two fire stations (\textit{S1} and \textit{S2}) responsible for rescuing fires in area \textit{A1}, and \textit{S1} had a significantly higher number of actions than \textit{S2}. }
    \label{fig:A1}
    \vspace{-3mm}
\end{figure}

\begin{figure}[h]
    \centering
    \includegraphics[width=0.5\textwidth]{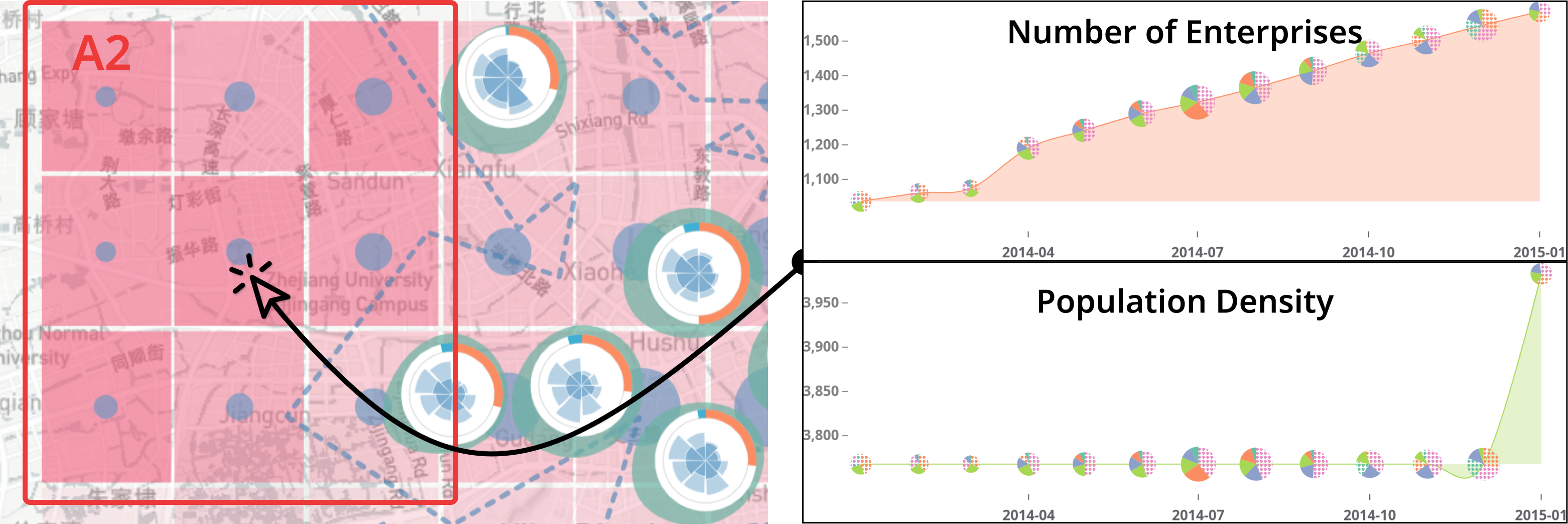}
    \vspace{-6mm}
    \caption{The information on the grid's characteristics showed that \textit{A2} was a densely populated and industrialized area.}
    \label{fig:A2}
    \vspace{-6mm}
\end{figure}

\par The experts aimed to investigate localized areas of imbalance in fire services within the city, in combination with reachable areas. Using the slider on the right, the map displayed the $10$-minute reachable areas from the fire station. Based on the color of the grid, size of the inner circle, and reachable areas, the experts identified two potential areas (\textit{A1} and \textit{A2}) that showed an imbalance between the supply and demand of firefighting resources. To begin their investigation, as shown in \cref{fig:A1}, E3 focused on \textit{A1}, which is situated far from the main city. He interacted with the map and identified two fire stations (\textit{S1} and \textit{S2}) responsible for attending fires in \textit{A1}. Comparing the outer circles of the glyphs, E1 discovered that \textit{S1} had significantly more actions than \textit{S2}. He then proceeded to hover the mouse over each of the two fire stations and found that \textit{S1} was assigned to rescue many fires closer to \textit{S2}. E1 analyzed that the resources were misallocated, stating, ``\textit{S1 shared too many tasks for S2, which caused S2 to be idle most of the time, while S1 exceeded the normal workload.}'' E2 concurred that this could be the cause of the long fire response time in \textit{A1}. The experts then proceeded to examine \textit{A2}. As depicted in the \cref{fig:A2}, the grid's characteristics showed that \textit{A2} was densely populated and industrialized. Upon further inspection of the geographic information, E3 observed that \textit{A2} primarily consists of residential neighborhoods and commercial areas, with a significant number of high-rise buildings. E3 noted that ``\textit{this building type usually has a more urgent need for fire protection services}''. However, due to the proximity to the central city, the workloads of the fire stations near \textit{A2} were relatively high, making it challenging to provide a rapid response to \textit{A2}. E1 observed, ``\textit{Unlike A1, which has spare firefighting resources available for deployment nearby. A2 is effectively facing a resource crunch.}'' Therefore, E1 recommended constructing one or two new fire stations in \textit{A2} to address this issue.

\par
\begin{figure}[h]
    \centering
     \vspace{-3mm}
    \includegraphics[width=0.5\textwidth]{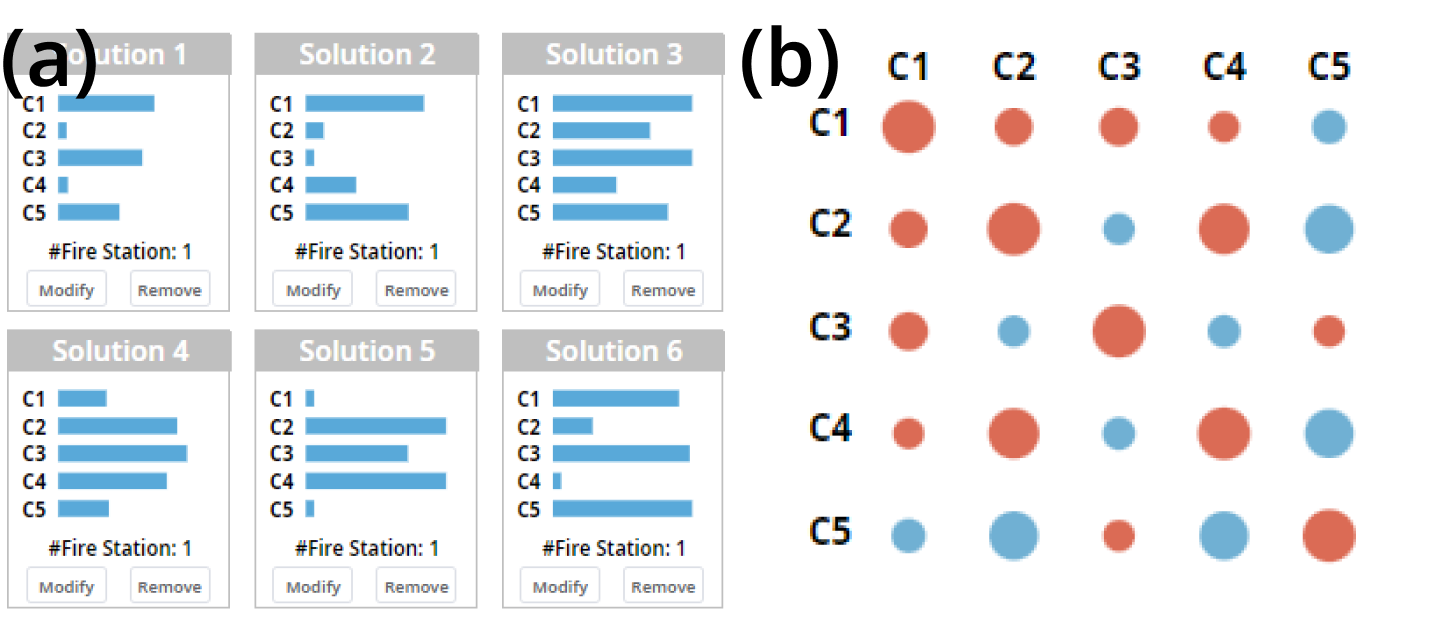}
    \vspace{-6mm}
    \caption{(a) Six candidate solutions. (b) Matrix plot representing the correlation between objectives.}
    \label{fig:sol}
    \vspace{-3mm}
\end{figure}

 Therefore, the group of experts shifted their focus towards the Optimization View. To begin, E1 directed his attention to \textit{A2}. He explained, ``\textit{it appeared that each objective had optimization value, and I want to assess the addition of a fire station based on these objectives first.}'' With C1 through C5 selected as optimization objectives and the number of new fire stations set at $1$, the model produced six solutions, as depicted in \cref{fig:sol}(a). E1 observed that Solution $\#1$, Solution $\#2$, and Solution $\#6$ were highly optimized for C2 and C4, but their optimization levels were relatively poor for the remaining objectives. E1 was curious as to why both C2 and C4 were simultaneously well optimized in these three solutions. The correlation between objectives can be viewed in the matrix plot (\cref{fig:sol}(b)). The red color represents a positive relationship between the objectives, indicating that they can be optimized concurrently. In contrast, the blue color indicates a negative relationship, meaning that optimizing one objective must come at the cost of the other. E3 explained that ``\textit{the correlation is also heavily influenced by the distribution of fires in the selected region.}'' After analyzing the solutions, E1 concluded that Solution $\#1$, Solution $\#2$, and Solution $\#5$ were superior choices as they displayed a high degree of optimization on two objectives, respectively. E1 then proceeded to examine the impact of the three solutions on the existing layout by utilizing the Simulation and Comparison View.

\begin{figure*}[h]
    \centering
        \vspace{-1mm}
    \includegraphics[width=\textwidth]{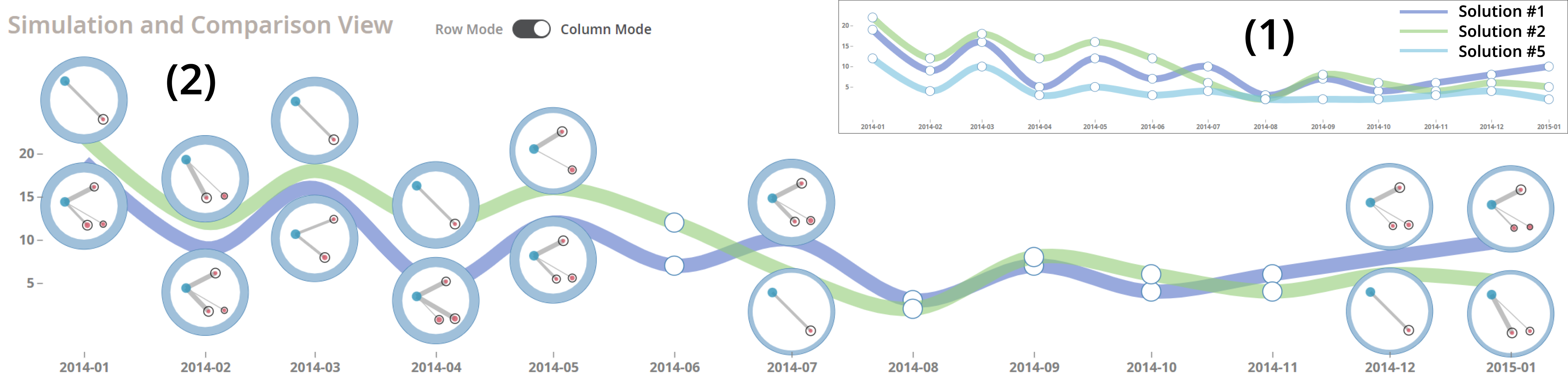}
    \vspace{-6mm}
    \caption{(1) Solution $\#1$ and Solution $\#2$ each have stronger priming ability at certain future time periods, while Solution $\#5$ is always weaker than them. (2) Solution $\#2$ delivers a higher number of transfers in the short term, its effectiveness in mitigating the regional response time diminishes over time.}
    \label{fig:sim}
    \vspace{-6mm}
\end{figure*}

\par E1 observed, from \cref{fig:sim}(1), that Solution $\#1$ and Solution $\#2$ possess varying degrees of ``transfer rescue'' ability at different future time periods, whereas Solution $\#5$ always exhibits weaker performance. Hence, E1 decided to eliminate Solution $\#5$ and focus on conducting a comparative analysis between Solution $\#1$ and Solution $\#2$ utilizing the column mode. As shown in \cref{fig:sim}(2), E1 discovered that although Solution $\#2$ delivers a higher number of transfers in the short term, its effectiveness in mitigating the regional response time decreases over time. Conversely, Solution $\#1$, while initially having a lower number of transfers, demonstrates better stability and continuity over time. E1 noted, ``\textit{Solution $\#2$ may have only resolved the emergency problem at the outset, but failed to sustain its advantage. In contrast, Solution $\#1$ is more adaptable to changes in the environment and fluctuations in demand, making it a more sustainable and stable strategy.}''

\subsection{Expert Interview}
\par To validate the efficacy of our approach, a semi-structured interview was conducted with E1-E3, lasting approximately one hour. Additionally, to ensure a comprehensive evaluation, we sought the involvement of two domain experts, namely E4 (a male urban facilities planner, aged 47) and E5 (a female researcher, aged 35), who were not engaged in the system design. These experts were invited to engage with and experience the functionalities of \textit{FSLens}, and their insights were incorporated through their active participation in the interview process.

\par \textbf{System Performance.} The proficiency of our system in supporting the interactive assessment and optimization of fire station layout received recognition from all experts. E1 expressed that our system greatly increased their operational efficiency by aiding in the identification and exploration of potential areas with limited firefighting resources. E1 further commented on the effectiveness of \textit{FSLens} in analyzing numerous fire events at varying levels of detail, which significantly decreased the workload required for traditional data exploration. In E1's words, ``\textit{I am very pleased with this.}'' E4 expressed a similar perspective, arguing that \textit{FSLens} improves the efficiency of planning decisions. E2 appreciated the ability of our visual design to effectively explain the algorithm-assisted decision-making process. Specifically, E2 praised that ``\textit{both feature importance and correlations between criteria are well embedded in the system, with intuitive designs that help understand the decision process.}'' The workflow and interaction of our system equally pleased the experts. E1 noted that the system's workflow aligned with the decision-making process in practice while adhering to the principle of moving from shallow to deep interactions. E3 was particularly impressed with our simulation approach and commented that ``\textit{with the Simulation and Comparison View, I can easily obtain the impact of optimization solutions over time, which provides valuable forward-looking insight.}'' Moreover, our user-centered design process facilitated the swift familiarization of experts with the visual coding employed. E5 expressed that ``\textit{although not being directly engaged in the initial design phase, the majority of the visual coding elements were intuitive and easy to understand}''. E4 supplemented this viewpoint, noting that ``\textit{while certain designs within the Spatiotemporal View exhibited a degree of complexity, the hierarchical organization was effectively executed}''. Furthermore, E4 emphasized that accessing specific details was facilitated through efficient filtering mechanisms.

\par \textbf{Visualization and Interaction.} According to the experts, our visualizations were deemed well-designed and user-friendly. E3 particularly commented on the Fire Service S\&D View, stating that ``\textit{it could effectively signal any significant disparities in firefighting resources in a timely manner.}'' E1 found the map design to be effective in ``\textit{identifying areas where firefighting resources may be lacking and exploring the underlying causes}''. The interactive workflow received commendations from both E2 and E5, who expressed their appreciation for its advanced technological integration, highlighting its ability to ``\textit{improve the efficiency and quality of data analysis through an interactive pipeline}''.

\section{Discussion and Limitation}
\par \textbf{Contributions Over Previous Work.} \textit{FSLens} differs from previous studies on interactive VA for site location selection in several ways. First, we begin by evaluating the supply and demand of firefighting resources, taking a ``holistic-to-local'' perspective to address regional imbalances and focus on specific localities. This approach is more practical and cost-effective compared to previous works that employed regional optimization methods~\cite{dong2018study,yu2012optimization}. Second, \textit{FSLens} considers the impact on the existing layout when determining the placement of new fire stations, a factor overlooked in previous VA studies for facility siting~\cite{li2020warehouse,liu2016smartadp,weng2018homefinder,karamshuk2013geo}. Third, \textit{FSLens} incorporates a spatiotemporal sequence forecasting model to predict fire occurrence patterns and utilizes model features to elucidate the spatiotemporal distribution of fires, which enables the exploration of fire occurrence patterns at both global and local grid scales, representing a novel endeavor.

\noindent\textbf{Generalizability and Scalability.} Through discussions with experts, we explored the applicability of various components of \textit{FSLens} in different scenarios, aiming to identify aspects that could be implemented directly and those requiring customization. The experts expressed that the system design is highly adaptable and has the potential to be utilized in other regional analyses of fire station layouts, given the availability of relevant data. Furthermore, one expert suggested the potential extension of the system to other urban public service scenarios, including medical rescue services, emergency rescue services, and utility services. Regarding scalability, \textit{FSLens} demonstrates its capability to handle a significant volume of fire rescue data, even reaching tens of thousands of records from a single city. This is achieved through the implementation of a grid-based division approach that organizes and presents the data hierarchically on a map. However, it's important to note that although the majority of the visual coding framework is scalable, challenges may arise when comparing multiple solutions simultaneously in the Simulation and Comparison View, leading to potential visual confusion. Moreover, the complexity of the algorithm used in \textit{FSLens} poses a bottleneck when dealing with multi-criteria optimization tasks. Specifically, the \textit{NSGA-\uppercase\expandafter{\romannumeral2}} algorithm requires a substantial population size and a high number of iterations to effectively converge, particularly in large target areas with numerous objectives. Unfortunately, this limitation results in a time-consuming process in such scenarios.

\noindent\textbf{Limitations.} First, historical records of early fire departments may contain inconsistencies and insufficiencies in fire data and standards across different regions, which could introduce biases in analyzing fire department rescue efficiency. Second, when evaluating the impact of new fire station placements on the original layout, the simulation of travel times relied on the current road network configuration, potentially disregarding historical context and real-time traffic conditions that can significantly influence travel times~\cite{li2020warehouse,liu2016smartadp}. Additionally, while \textit{FSLens} incorporates road network information, it does not support real-time analysis of road conditions. Third, the lack of relevant data on fire station characteristics, such as size, staffing, and equipment availability, means that \textit{FSLens} does not include operational capacity in the analysis, which could result in the underutilization of fire stations. Fourth, in the Spatiotemporal View, the fusion of reachable area boundaries of fire stations reduces visual clutter but makes it challenging to observe individual station reachable areas distinctly. Lastly, the evaluation of \textit{FSLens} lacks numerical comparisons due to the unavailability of publicly accessible firefighting datasets for benchmarking and the scarcity of comparable open-source systems.

\section{Conclusion and Future Work}
\par We introduce a novel visual analytics approach to assess and optimize the spatial arrangement of fire stations. It involves collaborating closely with domain experts to identify design requirements and introduces a visual analytics system called \textit{FSLens}. The system aids experts in examining and improving firefighting resources in potentially underserved urban areas based on fire distribution and existing fire station layout analysis. A case study validates the approach with expert feedback. In future investigations, we will incorporate additional factors such as fire causes and incident location types, aiming to provide a more comprehensive perspective on the underlying relationships between fire occurrence patterns and specific regions within urban areas, thus offering more forward-looking insights that can be leveraged to optimize fire station layouts. Additionally, integrating fire station operational capability measurements will enhance decision-making and overall spatial layout optimization effectiveness.

\acknowledgments{This work is partially supported by the Research and System Development of Resilient Urban Fire Risk and Fire Safety Dynamic Assessment Technology Based on Multiple Data Sources (Project Number: 2020XFZD19), Shanghai Frontiers Science Center of Human-centered Artificial Intelligence (ShangHAI), and Key Laboratory of Intelligent Perception and Human-Machine Collaboration (ShanghaiTech University), Ministry of Education.}

\balance
\bibliographystyle{abbrv-doi-hyperref-narrow}

\bibliography{template}
\end{document}